\documentclass[conference]{IEEEtran}
\IEEEoverridecommandlockouts
\usepackage{cite}
\usepackage{amsmath,amssymb,amsfonts,stackrel,eqnarray,}
\usepackage{algorithmic}
\usepackage{graphicx}
\usepackage{textcomp}
\usepackage{xcolor}
\usepackage{multirow}

\def\BibTeX{{\rm B\kern-.05em{\sc i\kern-.025em b}\kern-.08em
    T\kern-.1667em\lower.7ex\hbox{E}\kern-.125emX}}

\begin{document}

\title{Ambient backscatter communications using LTE cell specific reference signals}

\author{\IEEEauthorblockN{Kalle Ruttik\IEEEauthorrefmark{1}, Xiyu Wang\IEEEauthorrefmark{1}, Jingyi Liao\IEEEauthorrefmark{1},  Riku J\"{a}ntti\IEEEauthorrefmark{1}, and Phan-Huy Dinh-Thuy\IEEEauthorrefmark{2}}

\IEEEauthorblockA{\IEEEauthorrefmark{1}Department of Communications and Networking,
Aalto University,
02150 Espoo, Finland}
 Email: \{kalle.ruttik, xiyu.wang, jingyi.liao, riku.jantti\}@aalto.fi
 
 \IEEEauthorblockA{\IEEEauthorrefmark{2}Networks, Orange Innovation, Chantillon, France}
 Email: dinhthuy.phanhuy@orange.com
}

\maketitle

\begin{abstract}
Long Term Evolution (LTE) systems provide ubiquitous coverage for mobile communications, which makes it a promising candidate to be used as a signal source in the ambient backscatter communications. In this paper, we propose a system in which a backscatter device modulates the ambient LTE signal by changing its reflection coefficient and the receiver uses the LTE Cell Specific Reference Signals (CRS) to estimate the channel and demodulates the backscattered signal from the obtained channel impulse response estimates. We first outline the overall system, discuss the receiver operation, and then provide experimental evidence on the practicality of the proposed system.
\end{abstract}

\section{Introduction}
Ambient Backscatter Communications (AmBC) is a promising new technology to alleviate costs, power consumption, and spectrum occupancy. In AmBC, Backscatter Devices (BDs) can communicate with each other by modulating and reflecting received Radio Frequency (RF) signals from ambient sources. The concept was originally proposed in~\cite{liu2013ambient} which uses digital television broadcast signals as an ambient source. Other signals of opportunity include the FM broadcast signals~\cite{wang2017fm}, WiFi~\cite{bharadia2015backfi}, and cellular signals ~\cite{chi2020leveraging}. AmBC devices consume several orders of magnitude less power than their active transmitter based counterparts and are simpler and are cheaper to manufacture ~\cite{talla2017lora}. These features make them ideal for ultra-low power machine type communications \cite{duan2020ambient}.

The 6th generation (6G) of mobile networks is still at a research stage. However, the European flagship project, Hexa-X, already sets objectives of societal and environmental sustainability~\cite{6GZED}. \cite{6GZED} introduces a new type of mobile device in the network, harvesting ambient energy to power itself: the zero-energy device (ZED). Additionally, in some countries and cities, the deployment of new networks faces stringent Electromagnetic Field Exposure (EMFE) limits~\cite{Orange6GVision}. Recently, to provide an asset-tracking service in a sustainable and EMFE-aware manner, the new concept of “Crowd-Detectable ZED” (CD ZED) has been proposed~\cite{phan2021ambient}. A CD-ZED is a new type of RF tag, harvesting solar energy, backscattering waves from base stations to broadcast its identity, and detectable by any smartphone nearby, connected to the network. Thanks to an anonymous participation of the crowd of connected and geo-localised smartphones, a package stamped with a CD ZED is tracked almost “out-of-thin-air”~\cite{phan2021ambient}. This concept applies to all generations, from the 4th to the 6th.
Long Term Evolution (LTE) is an especially attractive source signal for CD ZED due to its ubiquitous coverage. Previous work on LTE signal based ambient backscatter communications~\cite{chi2020leveraging} use frequency shifting at the tag to move the backscattered signal away from the LTE band, which increases the bandwidth usage. In this work, we consider a low-data rate design in which the backscatter device shares the downlink bandwidth with the mobile system, and the channel estimator at the UE is utilized to decode the backscattered signal. If the backscatter symbol duration is longer than the UE's channel equalization window, then the UE channel estimator is able to track the channel variation caused by the BD. The device simply appears as an additional multipath component from the receiver's point of view. Hence, in that case, the BD can share the channel with downlink signal without causing harmful interference~\cite{ruttik2018does}. The use of a long symbol duration at the backscatter device side limits the data rate that these systems can use to some hundreds of bits per second. Even if the data rate is low, it would still enable the CD ZED.

Our proposed receiver uses the LTE primary and secondary synchronization to achieve radio frame, subframe, and symbol synchronicity with the base station transmitter, as well as to identify the center of the channel bandwidth and to deduce the Physical Channel Identity (PCI). After that, downlink Cell Specific Reference Symbols (CRS) are utilized for estimating the Channel State Information (CSI). In LTE downlink CRS is permanently present, even if there is no scheduled data, hence making the CD ZED data transmission independent of the cellular data transmission. We convert the LTE frequency domain channel estimated into time domain impulse response. In this paper, we focus on the case in which the relative delay difference between the direct path and BD modulated path is small, i.e., the BD is located near to the receiver. In this case, the BD information will be present in the first channel tap.

One of the key challenges in designing a receiver for in-band backscatter signals is the direct path interference. The strongest path from the base station to the user equipment is several orders of magnitude stronger than the backscatter modulated signal path. That is, the signal of interest is buried in the strong interfering signal. In this paper, we propose to separate the BD modulated signal from other scattered components in the frequency domain. We assume that the channel is changing slowly due to Doppler and that BD can use a symbol rate which is much larger than the maximum Doppler frequency in the channel. Consequently, we can separate the BD modulated signal from the multipath components by implementing a high-pass filter. To ensure a fast enough rate of change in the BD symbols, we apply Manchester encoding before transmission.

This paper is organized as follows: Section~\ref{se:LTEchannekl} discusses LTE channel estimation and how BD signal is visible at the obtained impulse response estimates; section~\ref{se:receiver} outlines the proposed receiver for the BD modulated signal; section~\ref{se:experimentation} describes the experimental setup utilized to verify the receiver design and finally section~\ref{conclusions} concludes the paper.

\section{LTE Channel estimation\label{se:LTEchannekl}}
\begin{figure}[t!]
\centering
\includegraphics[width=0.85\columnwidth]{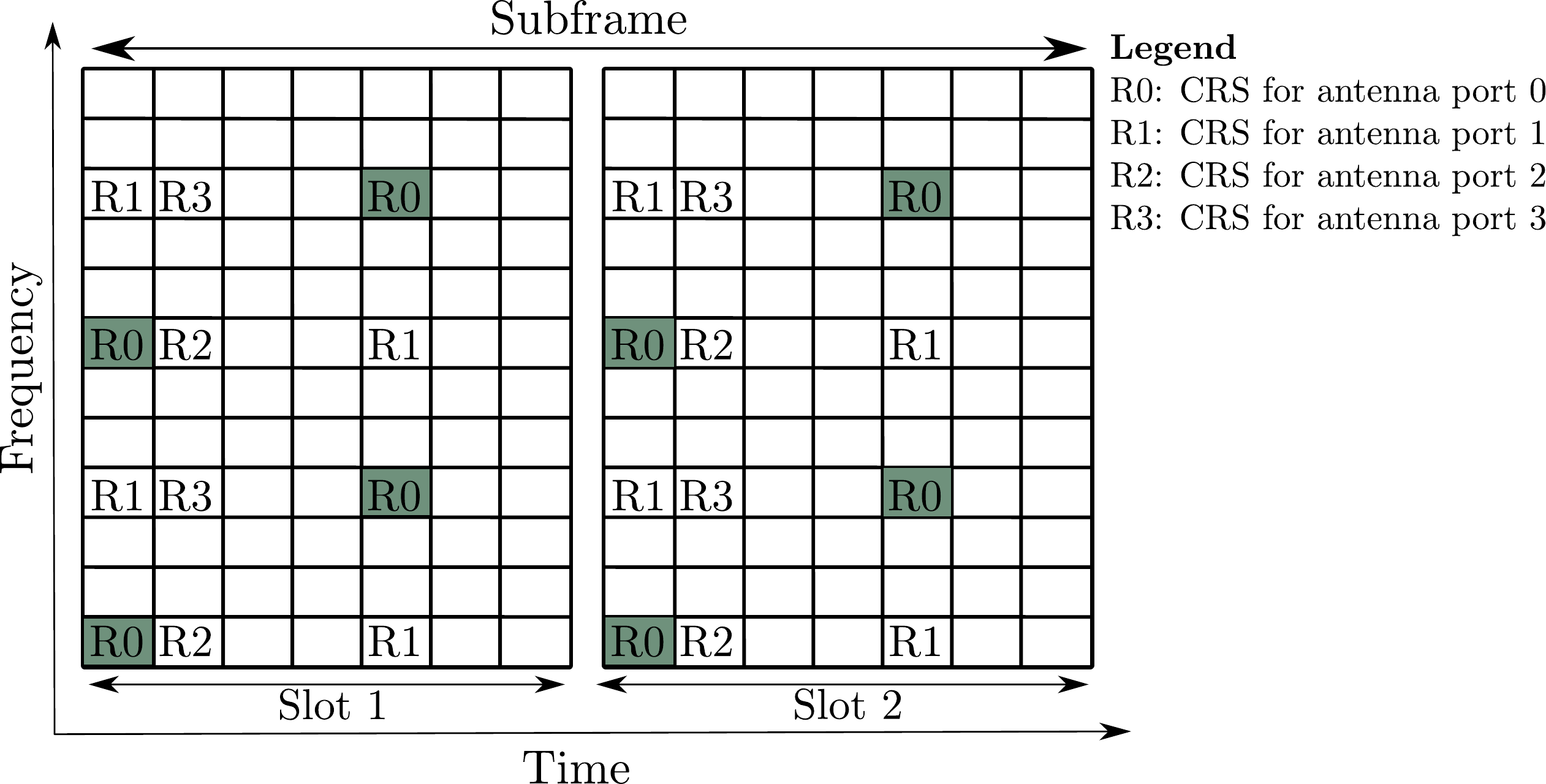}
\caption{ \scriptsize LTE Release 8 Cell Specific Reference Signal for antenna ports 0,1,2, and 3. 
  \label{fig:LTE_CRS}}
  \vskip -12pt
\end{figure}

In the LTE system, the base station transmits CRS in every subframe. Fig.~\ref{fig:LTE_CRS} illustrates the CRS reference signal allocation for antenna ports 0 to 3. LTE networks may operate using different CRS configurations. In a non-shifted CRS configuration, all cells use the same CRS time and frequency resources. In a shifted CRS configuration, different cells transmit CRSs on resources that are shifted in frequency. The non-shifted configuration avoids CRSs interference on data transmissions, but is also associated with a systematic CSI estimation error; especially noticeable at low traffic. Using the shifted configuration, the CRSs interfere with data transmissions, but the CSI estimation error is smaller \cite{madhugiri2014}. In this paper, we consider the case of shifted CRS, and the backscatter receiver uses pilots transmitted from antenna port 0.

The impulse response of the channel is
$$
h[\tau;t] = \sum_{k} a_k[t] \delta[\tau-\tau_k(t)] ,
$$
where $a_k(t)$, $\tau_k(t)$ are the time varying amplitude and delay of the $k^\textrm{th}$ multipath component and $\delta(\tau)$ denotes the Dirac's delta function. 
Let us assume that the channel stays constant at least during the transmission of a single LTE OFDM symbol of length $T_s=\frac{1}{W}$. Let us assume further that the path $k=1$ corresponds to the direct (shortest) path and the path $k=0$ corresponds to modulated scattered path. 
In that case $a_0[mT_s]=b_0[mT_s] x[m]$ where $x[m]$ denotes the  backscatter symbol and $b_0[mT_s]$ denotes the backscatter channel amplitude. We assume that the BD performs On-Off-Keying (OOK) such that $x[m]\in\{0,1\}$. 

The LTE receiver uses the CRS to get a frequency domain channel estimate $H[n;m]$, $m\in\{0,4,7,11\}$. For each OFDM symbol containing CRS, we can obtain the discrete time channel impulse response estimate by performing Inverse Fast Fourier Transform (IFFT).  Let $W$ denote the bandwidth of the LTE signal. The estimated $l^\textrm{th}$ channel tap can be written as
$$
\hat{h}[l;m] =\sum_k a_k^b\left(m T_s\right){\mathrm{sinc}}\left(l-\tau_k\left(m T_s\right)W\right) + z[m] ,
$$
where $m$ denotes the discrete (sample) time,
$
a_k^b[t]= a_k[t] e^{-i2\pi f_c \tau_k[t]} 
$
is the baseband equivalent channel tap, $f_c$ denotes the carrier frequency, and $z[m]$ denotes the estimation error. 

For simplicity, we assume that the delay difference between the direct path component and backscattered components are small such that they fall in the same tap $l_0$. This assumption holds when the BD is in the proximity of the LTE receiver. That is for $k=0,1$: ${l_0}-\frac{1}{2} \leq \tau_k W\leq {l_0}+\frac{1}{2}$. Hence, the estimate of channel impulse response of tap $l_0$ is  
\begin{equation}\label{eq:tap0}
\hat{h}[l_0;m] =g_0[m]x[m] + g_1[m] + z[m],
\end{equation}
where
$$
g_0[m]=\tilde{b}_0[mT_s]e^{-i2\pi f_c \tau_0[mT_s]},
$$
$$
g_1[m]={\sum_{k\geq 1} \tilde{a}_1[mT_s]e^{-i2\pi f_c \tau_k[mT_s]}},
$$
and $\tilde{a}_k[mT_s]=a_k[mT_s]\mathrm{sinc}\left(l_0-\tau_k[mT_s])W\right)$ and  $\tilde{b}_0[mT_s]=b_0[mT_s]\mathrm{sinc}\left(l_0-\tau_k[mT_s])W\right)$.
After obtaining the estimates of the tap for the OFDM symbols contining the pilot, we interpolate the channel for the rest of the OFDM symbols. For that we use low-pass filter with a cutoff frequency 400 kHz. Filtering interpolates the channel estimates from the obtained sequences in the symbols $\{\hat{h}[l_0;m], m\in\{0,4,7,11\}$ to the remaining symbols that did not have pilots, i.e., for the symbols $m\in\{1, 2,3,5,6,8,9,10,12,13\}$.  
After this we have samples of the channel available at rate 14 kHz.

\section{Backscatter receiver \label{se:receiver}}
\begin{figure}[t!]
  \includegraphics[width=\columnwidth]{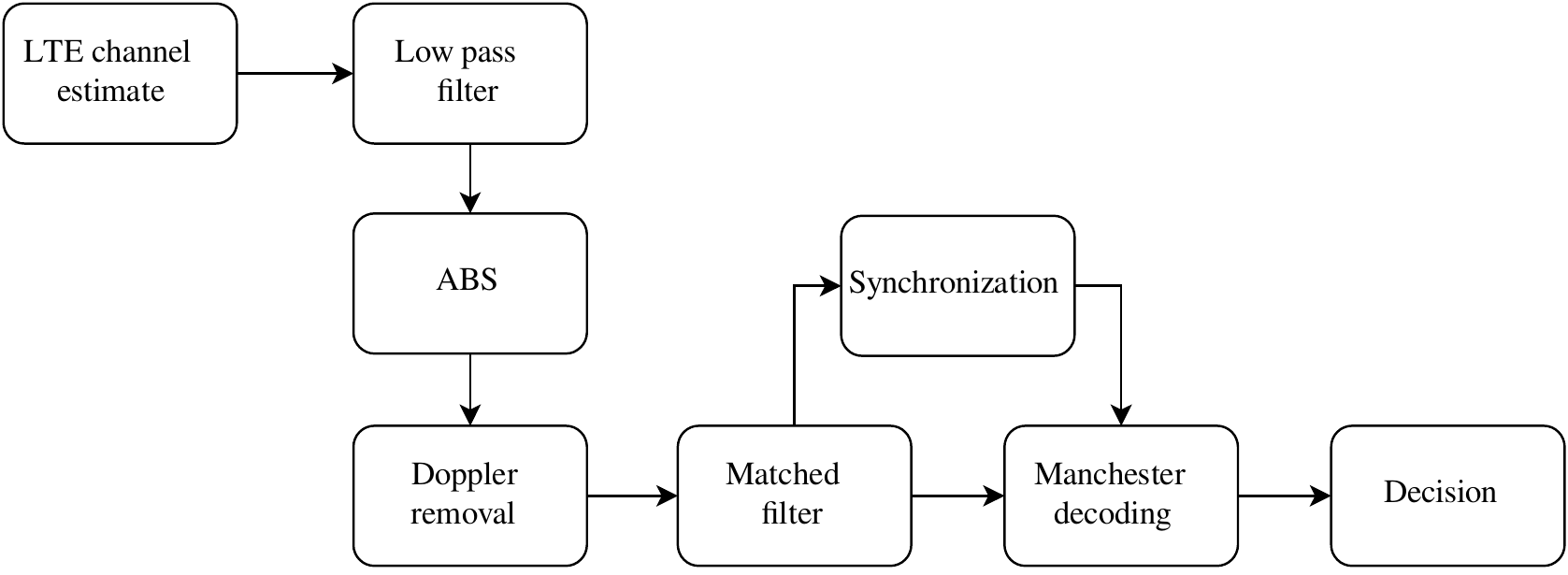}
  \caption{Flow chart of the proposed backscatter receiver.}
  \label{fig:Flowchart}
   \vskip -12pt
\end{figure}

The block diagram of the proposed Backscatter receiver is illustrated in Fig.~\ref{fig:Flowchart}. In what follows, we explain the operation of each block in more detail.

\subsection{Doppler removal}
The backscatter receiver needs to obtain the BD symbol and frame synchronicity and demodulate the BD signals. Prior to symbol synchronization, the receiver is not able to estimate the backscatter channel phase $\mathrm{arg}\{g_0[m]\}$. To get rid of the phase uncertainty, the receiver operates with the tap power $y[m]=|\hat{h}[l_0;m]|^2$. In a high LTE Signal-to-Noise Ratio (SNR) scenario, we have
$$
y[m]\approx   \alpha[m] +\beta[m]x[m] ,
$$
where $\alpha[m]\!=\!|g_1[m]|^2$ and $\beta[m]\!=\! (|g_0[m]|^2+2{\mathrm{Re}}\{g_1^*[m]g_0[m]\})$, and we consider the fact that $x^2[m] = x[m]$. The channels $\alpha[m]$ and $\beta[m]$ are changing in time because of the Doppler effect and possible frequency offset between transmitter and receiver. By changing the backscatter symbol at a higher frequency than the maximum Doppler in the channel, we can separate the direct path component $\alpha[m]$ and BD modulated path $\beta[m]$ in frequency domain.  In order to guarantee high variability in the BD message, we apply Manchester encoding for the BD symbols that map '0' to '(0,1)' and '1' to '(1,0)'. We then apply a high-pass filter on $y[m]$ to remove the direct path interference $\alpha[m]$. After the high pass filtering, the filtered signal $y_f[m]$ has zero mean and thus, the BD OOK symbols appear as binary phase shift keying (BPSK) modulated symbols.

\begin{figure}[t!]
    \centering
    \includegraphics[width=0.88\columnwidth]{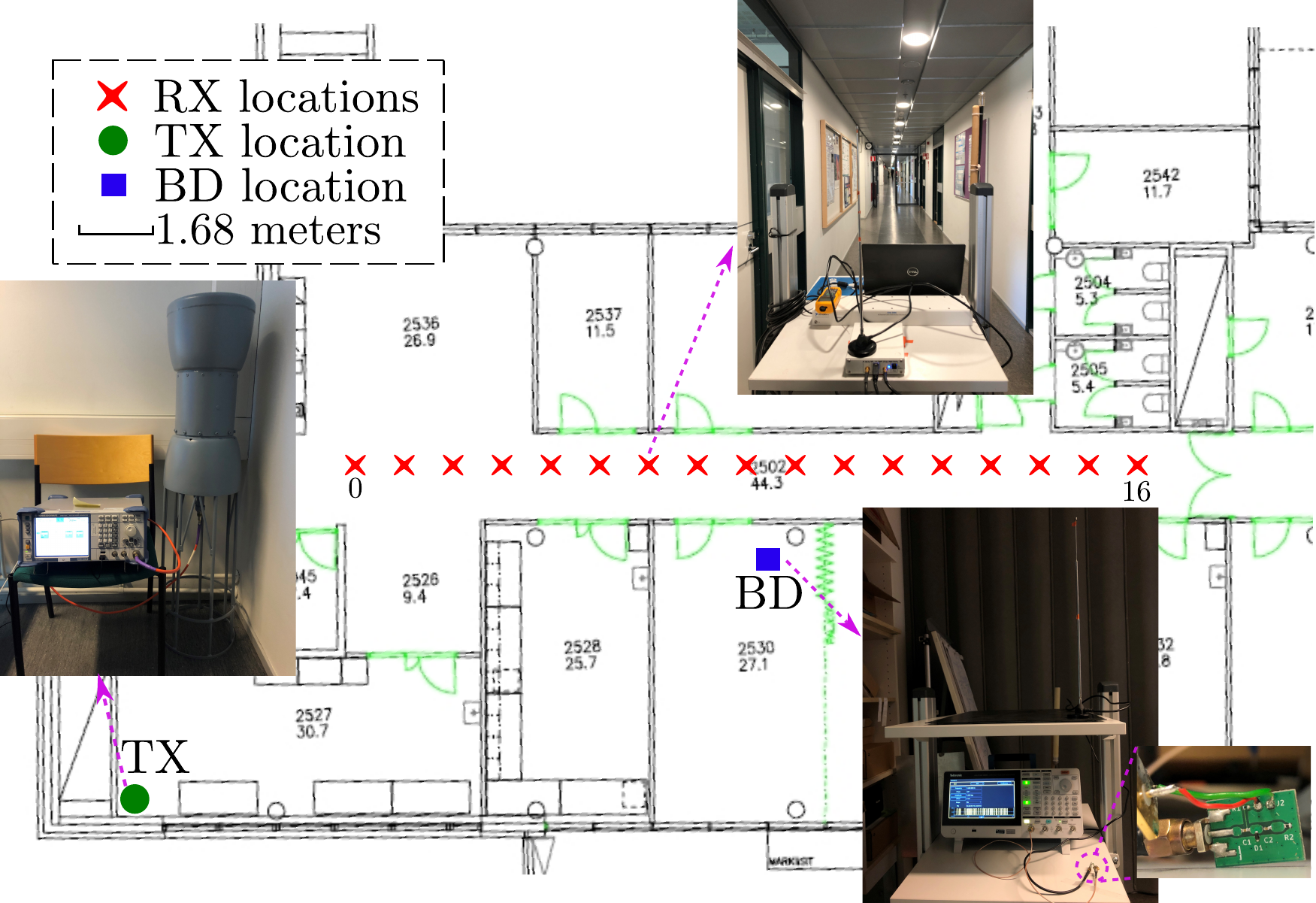}
    \caption{An illustration of the measurement setup.}
    \label{fig:corridor}
\vskip -12pt
\end{figure}

\subsection{Symbol and frame synchronization}
The BD frame consists of a synchronization sequence of two 13 bit Barker codes followed by 32 data bits. The data bits are Manchester encoded. The resulting packet consists 90 symbols. The symbol length is 10 ms. 

We consider that the BD symbol duration is longer than the OFDM symbol duration. Hence, we have oversampled the received filtered power signal $y_f[m]$. Before searching the packet start time we correlate with a matched filter to 10 ms symbols. After that, the packet start time is searched by correlating the matched filtered $y_f[m]$ with differently delayed versions of the synchronization header. The maximum correlation peak provides us with the start time of the frame.  We use the sign of the correlator output as an estimate of the channel phase and correct it accordingly. Considering the low autocorrelation sidelobe level ratio, two 13-bit Barker codes are applied on the synchronization in each frame.



\subsection{Demodulator}
We use correlator receiver to correlate the obtained data symbols with the Manchester encoded BPSK symbols $'(+1,-1)'$ and $'(-1,+1)'$. 

\begin{table}[t!]
\centering
\caption{Measurement setup parameters}
\resizebox{\columnwidth}{!}{
\begin{tabular}{l|c|c} \hline
   &Parameter &Value \\
   \hline
   \multirow{7}{*}{Tx} & \multirow{2}{*}{Baseband signal generator} &\multirow{2}{12em}{R\&S SMBV100A Vector Signal Generator} \\
   & &\\ \cline{2-3} 
   & Antenna & R\&S HK033 VHF/UHF coaxial dipole\\  \cline{2-3} 
   & Carrier frequency &486MHz \\ \cline{2-3} 
   & Bandwidth & 7.68 MHz \\ \cline{2-3} 
   & TX power level & 15 dBm \\ \cline{2-3} 
   & Peak envelop power &29.09 dBm\\
   \hline
   \multirow{7}{*}{BD}
   &\multirow{2}{*}{Baseband signal generator} & \multirow{2}{12em}{Tektronix AFG 31000 Arbitrary Function Generator} \\ & &\\ \cline{2-3} 
   & Antenna & RaTLSnake M6 telescopic antenna\\  \cline{2-3} 
   & Symbol duration & 10 ms \\ \cline{2-3} 
   & Synchronization & 13 bit Braker code \\ \cline{2-3} 
   & Encoding & Manchester \\ \cline{2-3} 
   & Modulation scheme & OOK\\ 
   \hline
   \multirow{3}{*}{Rx} & Device & NI USRP-B210 \\ \cline{2-3} 
   &Antenna & RaTLSnake M6 telescopic antenna\\ \cline{2-3} 
   &AD converter& 12 bits\\ \cline{2-3}
   \hline
\end{tabular}}
\label{tab:parameters}
\end{table}

\section{Experimental results \label{se:experimentation}}
To verify the proposed receiver, we conduct experiments inside Maarintie 8 building on the Aalto University campus. Fig.~\ref{fig:corridor} illustrates the  measurement setup. We place the LTE transmitter (TX) and the BD in two different offices. We select 17 positions for the RX in the corridor. Every two adjacent points are 1 meter apart and the points from left to right are respectively marked as 0 to 16. By using the proposed algorithm, we test the coverage area of the BD in the indoor environment. We use Rohde \& Schwarz (R\&S) SMBV 100A signal generator to send LTE signal having 7.68 MHz bandwidth and 486 MHz carrier frequency with an average transmit power of 15 dBm. As a TX antenna, we use R\&S HK300 antenna. The BD modulator is driven from Tektronix AFG 31000 signal generator

At the RX end, we utilize the National Instrument (NI) Universal Software Radio Peripheral (USRP)-B210 with a single omni-directional antenna. The USRP is connected to a laptop for acquiring and post-processing the data.
The detailed configuration parameters are shown in Table~\ref{tab:parameters}.

The measurements were conducted without automatic gain control. The B210 has only 12 bit AD converter. In order to guarantee sufficient dynamic range at each measurement point, the USRP gain was hand tuned to be the maximum possible without clipping the received signal. Compared with direct signal, BD signal is too weak to be detected under quantization noise. At each measurement location we sent 4 second of the signal. Three times of measurements are repeated at each location to decrease measurement error. The signals were post-processed in Matlab by using the LTE toolbox.

\begin{figure}[t!]
\centering
\includegraphics[width=0.82\columnwidth]{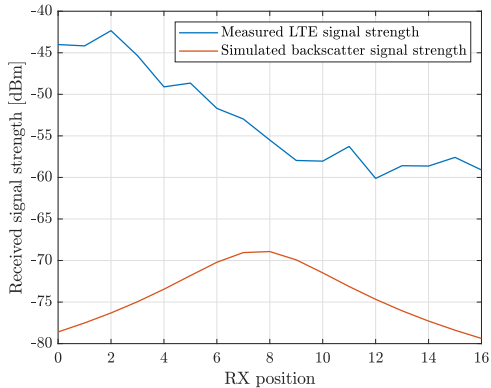}
\caption{Variation in the measured received LTE signal power and the simulated backscatter signal power at the 17 RX positions. The backscatter signal power is calculated using the free space path loss with exponent of 2. \label{fig:power} }
\vskip -12pt
\end{figure}
\begin{figure}[t!]
\centering
    \includegraphics[width=0.82\columnwidth]{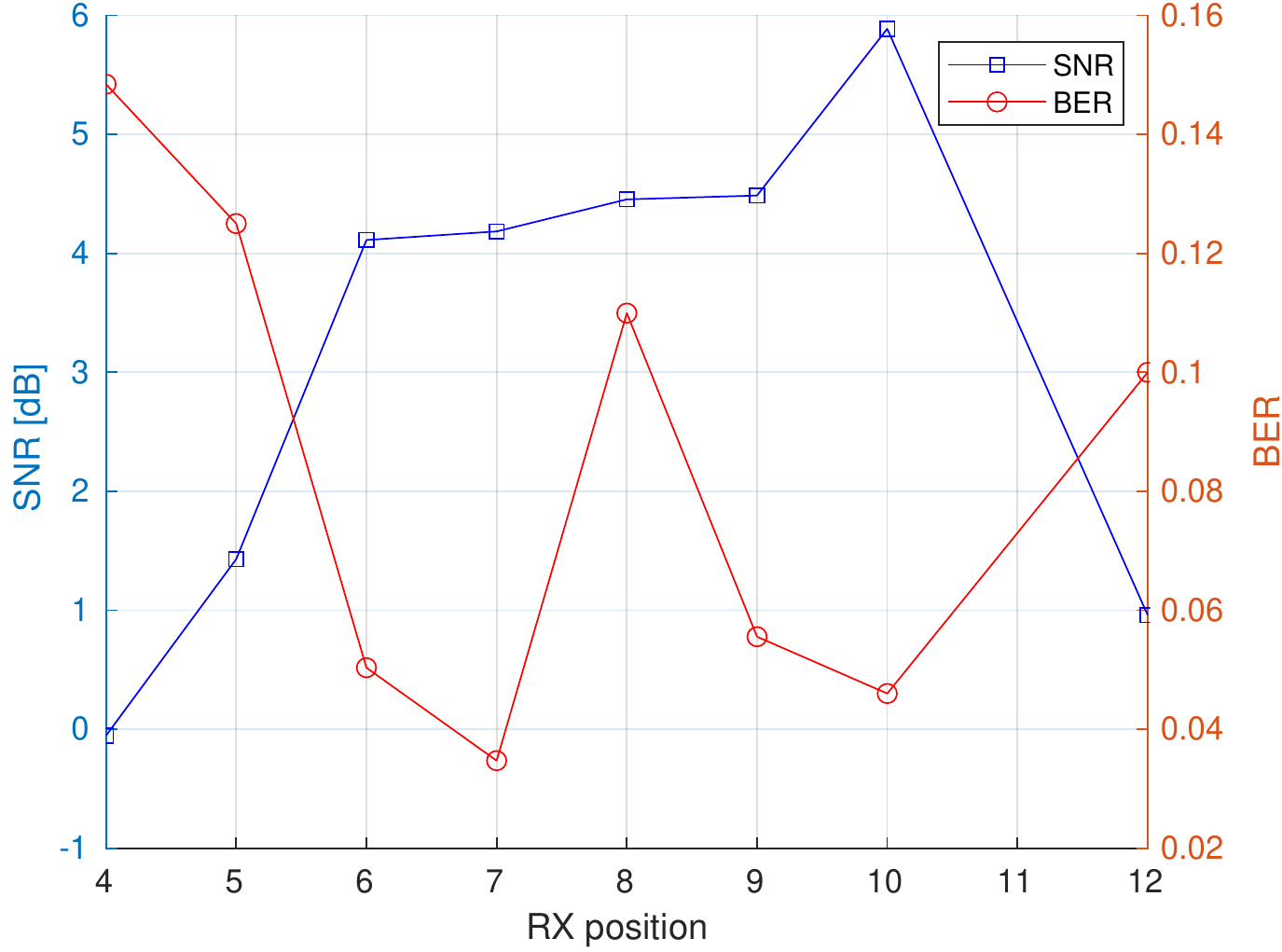}
  \caption{Measured SNR and BER at the positions where receiver could detect the signal. \label{fig:SNR_and_BER} }
  \label{fig:ReceivedSignal}
\vskip -12pt
\end{figure}

%
Fig.~\ref{fig:power} illustrates the measured LTE CRS power and calculated theoretical backscatter signal power based on the model \cite{griffin2009complete} at the measured positions. The LTE signal level was measured by Tektronix RSA6114A spectrum analyzer.  Even though the considered environment is much lossy, the simulated optimistic estimate of the backscatter signal power is included for the illustrative purpose. 
%
As can be seen from the figure, the backscattered signal power is much weaker than the direct path power, and it increases when the BD is close to the RX. Fig.~\ref{fig:SNR_and_BER} illustrates the measured backscatter signal SNR and Bit Error Rate (BER) at the different locations. The results are present only for the locations where the receiver was able to successfully synchronize to the BD frame and decode the data. The BD symbol rate was fixed to 100 bit/s, and we recorded three packets. The backscatter SNR varied from 0 to 6 dB, and the obtained raw BER varied between 0.15 to 0.04 based on the location. The BER is computed over relatively few packets. For better estimation and improving the receiver performance, a larger measurement campaign will be carried out in the future. However, this proof of concept implementation indicates the feasibility of the backscatter communication. The connection can further be improved with better channel estimation, equalization and by adding forward error correction. 



\section{Conclusions \label{conclusions}}
In this paper, we proposed a system that uses the LTE cell specific reference signals and channel estimator for receiving backscatter modulated signals. Due to the space limitation, we only outlined the receiver structure and discussed the case where the backscatter device was in the proximity of the receiver. The proposed receiver was validated using over the air measurements in an indoor environment. Based on our experimental results, we can conclude that the LTE channel estimator offers a great potential to be utilized for receiving backscattered signals in CD ZED applications.

\section{Acknowledgements}
This work is in part supported by the European Project Hexa-X under grant 101015956 and Academy of Finland ULTRA and BESMIAL project under grants 328215 and 334197, respectively.
\newpage
~
\vskip 11pt

\bibliographystyle{IEEEtran}
\bibliography{References}

\end{document}